\documentstyle[twoside,fleqn,espcrc2,epsf]{article}

\title{Improving the staggered quark action to reduce
       flavour symmetry violations.%
       \thanks{Talk presented by D.~K.~Sinclair, at LATTICE'97, Edinburgh,
               Scotland.}}

\author{J.-F.~Laga\"{e} and D.~K.~Sinclair%
       \address{HEP Division, Argonne National Laboratory, 9700 South Cass
                Avenue, Argonne, IL, 60439, USA}%
       \thanks{Supported by DOE contract W-31-109-ENG-38.}}
        
\begin{document}
  
\begin{abstract}
We investigate a class of actions for lattice QCD with staggered quarks
aimed at reducing the flavour symmetry violations associated with using
staggered fermions. These actions replace the gauge field link fields in
the quark action with covariantly smeared fields. As such they are an
extension of actions considered by the MILC collaboration. We show that
such actions systematically reduce flavour symmetry violations in the weak
coupling limit. Using the mass splitting between Goldstone and non-Goldstone 
pions as a measure of flavour symmetry violations we find that these actions
have considerably less flavour symmetry violations than the standard staggered
action, and represent an improvement on what can be achieved with the MILC
action, on quenched configurations with $\beta=5.7$.
\end{abstract}
\maketitle

\section{INTRODUCTION}

Improving the lattice QCD action using the method of Lepage et al.
\cite{lepage} and the perfect action method \cite{hn} has allowed simulations
with relatively large lattice spacings ($\sim 0.5fm$). While these methods have
proved successful for lattice QCD with Wilson \cite{lepage} or non-relativistic
quarks \cite{nrqcd} they have been less successful for staggered
(Kogut-Susskind) quarks \cite{milcspec,peikert}.

Improved staggered quark actions such as the Naik(1+3-link) \cite{naik}, aimed
at improving the fermion dispersion relations to bring them closer to their
continuum form, do provide some improvement. However, they do not significantly
decrease the flavour symmetry violations which plague the staggered quark
action. The simplest of ``perfect'' actions also exhibit this deficiency
\cite{peikert}. 

The reason appears to be that these actions concentrate on improving
the low momentum dispersion relations for the fermions associated with the
``corners'' of the Brillouin zone (and the gauge fields) and on making the
interactions between low momentum fields more continuum-like. Flavour symmetry
violation occurs because the momentum transferred to the fermions by the gauge
fields can change them from one flavour to another, and such transfers are
not suppressed by such improvements.

Recently the MILC \cite{milc} collaboration have introduced an action where the
gauge links in the staggered fermion action are replaced with a linear
combination of a single gauge link and a sum over the six 3-link ``staples''
joining the same sites. This does appreciably decrease flavour symmetry
violations as measured by the splitting between the Goldstone and non-Goldstone
pions. The reason for this is clear. Such a replacement effectively smooths the
gauge field interaction with the fermions, thus reducing large momentum
transfers at the vertex responsible for flavour symmetry breaking. 

We have improved on this scheme in a systematic fashion. At tree level, our
scheme suppresses flavour symmetry breaking by an additional power of $a^2$.
We assume that an action of the same form, but with different coefficients, is
the appropriate choice beyond tree level.

\section{IMPROVED STAGGERED-QUARK ACTION}

The 4 flavours and 4 Dirac components in one staggered fermion are associated
with the 16 poles of the free propagator. For massless quarks these occur at
$p_\mu = 0, \pi$. Hence, at weak coupling, we can suppress flavour mixing to
higher order in $a$ if we can suppress coupling to gluons which couple the
neighbourhood of one of these poles to any other. In particular we should
suppress coupling to gluons with momentum components $0$ or $\pi$, but not all
zero.

To see how to do this let us ignore, for the moment, the complications of
gauge invariance. Then the coupling of the gluon and quark fields would be
of the form
\begin{equation}
i \psi^{\dagger}(n) \eta_\mu(n) A_\mu(n) \psi(n+\mu) + h.c..
\end{equation}
We could then achieve our objective by making the substitution
\begin{eqnarray}
\lefteqn{A_\mu(n) \rightarrow \frac{1}{256}(2+D_1+D_{-1})(2+D_2+D_{-2})} 
\nonumber \\
       &  & \times (2+D_3+D_{-3})(2+D_4+D_{-4}) A_\mu(n)
\label{eqn:naive}
\end{eqnarray} 
where
\begin{equation}
D_{\pm\nu} A_\mu(n) = A_\mu(n\pm\nu)
\end{equation}
Then in momentum space
\begin{eqnarray}
A_\mu(k) & \rightarrow &\frac{1}{16}(1+\cos k_1)(1+\cos k_2) \nonumber \\
         &             &            (1+\cos k_3)(1+\cos k_4) A_\mu(k)
\end{eqnarray}
which vanishes when any component of $k$ is $\pi$, and $\rightarrow A_\mu(k)$
as $k \rightarrow 0$. For $k$ having any component within ${\cal O}(a)$ of 
$\pi$, the coupling is suppressed by $\sim a^2$. Thus this substitution would
achieve precisely what we want. Note also that adding a fermion mass shifts the
poles in the fermion propagator by ${\cal O}(a)$, and so does not invalidate
these arguments.

To make this approach gauge invariant, we replace $A_\mu$ by the standard gauge
link variable $U_\mu = \exp i A_\mu$. The displacement operators $D_{\pm\nu}$
are replaced by covariant displacement operators, for example
\begin{equation}
D_{\nu} U_\mu(n) = U_\nu(n) U_\mu(n+\nu) U^{\dagger}_\nu(n+\mu).
\end{equation}
As one moves away from weak coupling, the coefficients in this new action will
develop $g$ dependence, so that these coefficients are considered as free
parameters to be adjusted to minimize flavour symmetry violations.
Equation~\ref{eqn:naive} then becomes 
\begin{eqnarray}
\lefteqn{U_\mu \rightarrow {\bf U}_\mu  =  C\{x_0 + 2y_0}        \nonumber  \\
&& \hspace{-.1in} + \sum_\nu [x_1 D_\nu + y_1 (D_{\mu\nu}+D_{-\mu\nu})]   
                                                                 \nonumber  \\
&& \hspace{-.1in} + \sum_{\nu\rho} [x_2 D_{\nu\rho} + y_2 (D_{\mu\nu\rho} 
                                           + D_{-\mu\nu\rho})]   \nonumber  \\
&& \hspace{-.1in} + \sum_{\nu\rho\lambda} [x_3 D_{\nu\rho\lambda}
   + y_3 (D_{\mu\nu\rho\lambda}+D_{-\mu\nu\rho\lambda})]\} U_\mu 
\end{eqnarray}
where $\nu$, $\rho$ and $\lambda$ are summed over $\pm 1$, $\pm 2$, $\pm 3$,
$\pm 4$ with $|\mu|$, $|\nu|$, $|\rho|$, $|\lambda|$ all different, and 
$ C = 1/( x_0 + 2 y_0 + 6 x_1 + 12 y_1 +  12 x_2 + 24 y_2 + 8 x_3 + 16 y_3 ) $.
The $D$'s acting on $U_\mu$ displace it in each of the subscript directions. 
Products of $U$'s join the ends of the displaced link to the sites at the
ends of the undisplaced $U_\mu$ so as to ensure gauge invariance. To reduce the
number of independent parameters, we symmetrize over all such minimal length
paths joining the undisplaced sites to displaced sites. In general, we could
assign different weights to such paths, provided we preserved lattice 
symmetries.

The tree level values for these coefficients are $x_0=1$, $y_0=x_1=1/2$,
$y_1=x_2=1/4$, $y_2=x_3=1/8$, $y_3=1/16$. The MILC action has $x_0=1$,
$x_1=\omega$, and $x_2=x_3=y_1=y_2=y_3=0$. 

The first check of this fermion action is to see that it does indeed
suppress coupling to gluons with momentum components $0$ or $\pi$ but not all
zero, in the weak coupling limit when the coefficients are set to their tree
values. For this we use the weak coupling approximation
\begin{equation}
U_\mu \approx 1 + iA_\mu .
\end{equation}
and expand ${\bf U}_\mu$ to first order in $A_\mu$. Since we are dealing with
non-gauge-invariant quantities, we work in Landau gauge. In this approximation
${\bf U}_\mu (k)-1$ vanishes at $k_\mu = 0, \pi$, except when 
$k_1=k_2=k_3=k_4=0$, which is the required result.

\section{TESTS OF NEW ACTION AT FINITE COUPLING.}

The true test of this new action is how it reduces the flavour symmetry
violation, as measured by the $\pi_2$ -- $\pi$ mass difference for values of
$\beta=6/g^2$ where this splitting is quite large for the standard action
\cite{milc}.

We have measured the spectrum of local light hadrons on an ensemble of 203
quenched configurations generated with the standard action at $\beta=5.7$ on
an $8^3 \times 32$ lattice. We have used the standard staggered quark action,
the MILC action and our new improved action, for various values of their
parameters. For the results presented here, we used a 1 parameter subset
of improved actions for which $x_n=x^n$, $y_n=x^{n+1}$. For the MILC actions
$x_0=1$, $x_1=\omega$, and $x_2=x_3=y_1=y_2=y_3=0$, as mentioned above.

We plot the standard action results for $m_\pi$ and $m_{\pi_2}$ in figure~1,
along with those for MILC1 (the MILC action with $\omega=1$) and imp2 (our
improved action with $x=1$), which show the most improvement of those
actions we have considered in each class. 
\begin{figure}[htb]
\vspace{-0.1in}
\epsfxsize=3in
\epsffile{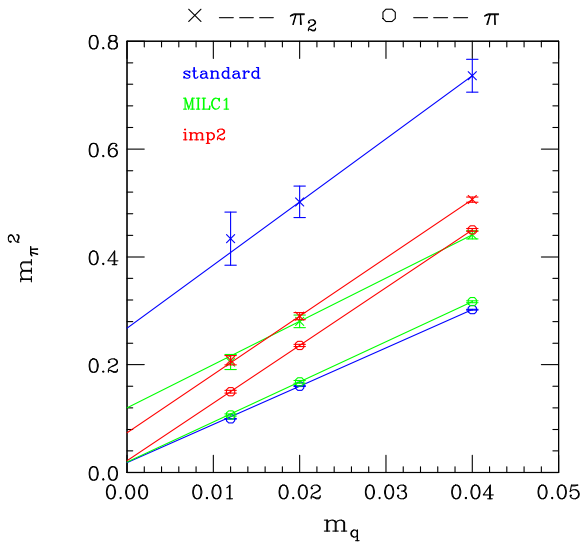}
\vspace{-0.1in}
\caption{Comparison of standard, MILC and improved actions for masses of the
         Goldstone pion, $\pi$ and local non-Goldstone pion, $\pi_2$.}
\end{figure}

Since the $\rho$ mass for the standard action, extrapolated to $m_q=0$, is
$\sim 15\%$ larger than that for the MILC or improved actions, we should
really compare these with standard action results from a lattice with
$\sim 15\%$ larger lattice spacing, where the improvement will be less dramatic.
However, a reduction of $\pi$--$\pi_2$ mass splitting of a factor of 2 for
the MILC actions, as reported at $\beta=5.85$ by the MILC collaboration seems 
likely.

Since the $\rho$ masses for the MILC and improved actions are very close, a
direct comparison of $\pi$--$\pi_2$ mass splittings is justified. Looking at
an extrapolation to zero quark mass suggests that our improved action could 
reduce this mass splitting by around 40\%.

\section{SUMMARY AND CONCLUSIONS}

Actions which are designed to systematically suppress the coupling of
high-momentum gluons, which give rise to flavour symmetry violations, reduce
flavour symmetry violations in the pion mass spectrum by $\sim 40\%$
over the MILC action, and hence by a factor of $\sim 3$ over the
normal staggered fermion action, on quenched configurations at $\beta=5.7$. 
We are repeating these calculations on a larger lattice ($16^3 \times 32$) to
remove finite size effects.

Such improvements should be combined with an improved gauge action, and with
a fermion action designed to improve the low-momentum dispersion relations
(Naik or perfect action).

More careful studies of such actions away from the weak coupling limit are
needed. Finally we should relate these actions to those studied by Luo
\cite{luo}.

\end{document}